\documentclass[aps,pre,amsmath,amssymb,twocolumn]{revtex4}      

\usepackage{graphicx}

\begin{document}

\title{Saddles and softness in simple model liquids}

\author{L.~Angelani}
\affiliation{
Dipartimento di Fisica, INFM-CRS SMC, 
Universit\`a di Roma {\em La  Sapienza}, P.le A. Moro 2, 00185 Roma, Italy
}
\author{C.~De Michele, G.~Ruocco, and F.~Sciortino}
\affiliation{
Dipartimento di Fisica, INFM-CRS Soft, 
Universit\`a di Roma {\em La Sapienza}, P.le A. Moro 2,  
00185 Roma, Italy
}
\date{\today}
\begin{abstract}
We report a numerical study of saddles properties of the potential energy
landscape for soft spheres with different {\it softness}, i.e. different
power $n$ of the interparticle repulsive potential.
We find that saddle-based quantities rescale into master curves once
energies and temperatures are scaled by mode-coupling temperature $T_{MCT}$,
confirming and generalizing previous findings obtained for Lennard-Jones 
like models.

\end{abstract}
\maketitle

The analysis of the potential energy landscape (PEL)
of model liquids has allowed to clarify many interesting phenomena
of the supercooled liquid regime and 
the slowing down of the dynamics 
\cite{deb_nature,inm,nature1,heuer,pel_th,donati,sad_ag,pel_ag}.
More recently a promising PEL description was obtained studying the properties
of saddle points \cite{noi_sad,cavagna,chakra,deb_prl}.
In a previous work \cite{jcp_ang1} some of us reported a numerical 
investigation of the PEL for different 
Lennard-Jones (LJ) like model liquids, 
focusing on the properties of saddle points
visited by the systems at different temperatures.
The main findings of that work were:
{\it i)} the existence of master curves for saddle-based quantities 
(saddle order $n_s$ vs. $T$, energy elevation of saddles from underlying 
minima vs. $n_s$) when temperatures and energies are scaled  by 
mode-coupling temperature $T_{MCT}$ (we set $k_B$=$1$);
{\it ii)} a nearly constant ratio between elementary saddle energy-barriers 
$\Delta E$ (energy elevation of saddles of order 1 from underlying minima)
and $T_{MCT}$: $\Delta E \simeq 10\;T_{MCT}$;
{\it iii)} a quantitative relationship between $\Delta E$
and the Arrhenius activation energy $\Delta E_{Arr}$ (obtained from low-T 
diffusivity): $\Delta E_{Arr} \simeq 2\;\Delta E$.
Although obtained for different models, the reported universality 
was not too surprising
due to the similar shape of the repulsive part of the pair potential,
in particular all having the same $r^{-12}$ dependence.
For this reason a wider class of models must be analyzed
in order to show the robustness of the reported universality.

In this note we extend the class of model liquids under consideration,
analyzing soft spheres with different
power $n$ of the interparticle repulsive potential (different {\it softness})
\cite{demic}.
We find that systems interacting with $r^{-n}$ potential belong to the same 
universality class of LJ-like potentials, thus pointing towards a common
organization of saddles in the PEL of disordered systems.
 
The investigated systems are $80$:$20$ binary mixtures of $N$=$1000$ 
particles enclosed in a cubic box with periodic boundary conditions and 
interacting through the pair potential
\begin{equation}
V_{\alpha \beta}(r) = 4 \epsilon_{\alpha \beta} 
\left( \frac{\sigma_{\alpha \beta}}{r} \right)^n \ , 
\end{equation}
where $\alpha,\beta \in A,B$, 
$\sigma_{AA}=1.0$, $\sigma_{AB}=0.8$, $\sigma_{BB}=0.88$, 
$\epsilon_{AA}=1.0$, $\epsilon_{AB}=1.5$, $\epsilon_{BB}=0.5$ \cite{note,bmlj}.
Reduced units will be used in the following ($\sigma_{AA}$ for length,
$\epsilon_{AA}$ for energy, $(m\sigma_{AA}^2/\epsilon_{AA})^{1/2}$ for time
- m is the mass of the particles).
The analyzed density was $\rho$=$1.2$.
The investigated values of the parameter $n$ tuning the {\it softness} of the 
interaction were $n$=$6,8,12,18$. We note that for $n<6$ crystallization
events prevent the study of the supercooled regime.
Following previous works \cite{noi_sad}, 
we studied the properties of the saddles of the PEL
visited by the system during its dynamic evolution at a given temperature
(performed through isothermal molecular dynamics simulations with 
Nos\'e-Hoover thermostat).
The temperature range investigated is such that, for each $n$,  the diffusivity
covers about 4 orders of magnitude.
Saddles are located using minimization procedures 
(LBFGS algorithm as implemented by Liu and Nocedal \cite{lbfgs})
on the pseudo-potential $W$=$|\nabla V|^2$ ($V$ is the total potential energy).
Inherent structures are also located (minimizing the true potential $V$) 
at each temperature. 
We then collected the properties of saddles (energy $e_s$ and order
$n_s$, defined as the number of negative eigenvalues of the Hessian of $V$)
and inherent structures (energy $e_{_{IS}}$).

\begin{figure}[tb]
\includegraphics[width=.47\textwidth]{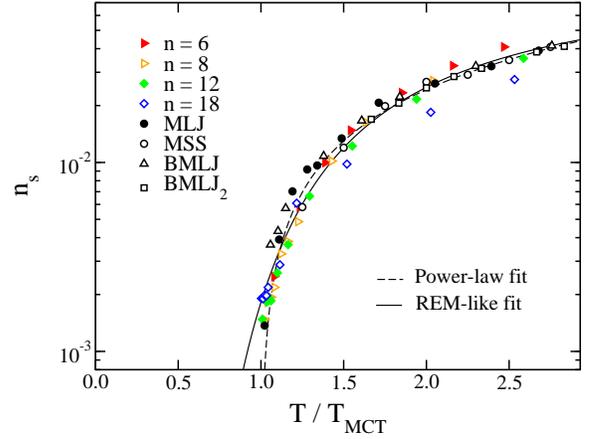}
\caption{
Saddle order $n_s$ as a function of
$T/T_{MCT}$ for the model systems of this work
(soft spheres with different {\it softness} $n$)
together with the data obtained in a previous work for LJ-like models \cite{jcp_ang1}.
Dashed line is a power-law fit, while full line is a REM-like fit \cite{REM_note}.
}
\label{fig1}
\end{figure}

In Fig. 1 we show the saddle order $n_s$ as a function of $T/T_{MCT}$ 
($T_{MCT}$ is the mode-coupling temperature estimated from the power law 
divergence of inverse diffusivity) 
for different values of {\it softness} $n$.
In the figure we report also the same quantity for the LJ-like systems 
analyzed in a previous work \cite{jcp_ang1}. 
All the data collapse onto a master curve,
even if some deviation is present, in particular for $n$=$6$ and $n$=$18$.
In a first approximation scaling $T$
by $T_{MCT}$ gives rise to a common behavior of the analyzed quantity
for the different models.
It is worth noting that a similar master plot has been obtained 
in Ref. \cite{REM_keyes} for LJ and CS$_2$ systems.

Looking at the mean energy elevation $e_s - e_{_{IS}}$ of saddles 
from underlying minima as a function of saddle order $n_s$ 
(inset in Fig. 2),
one obtains nearly straight lines, with different slopes for different 
values of $n$.
\begin{figure}[t]
\includegraphics[width=.47\textwidth]{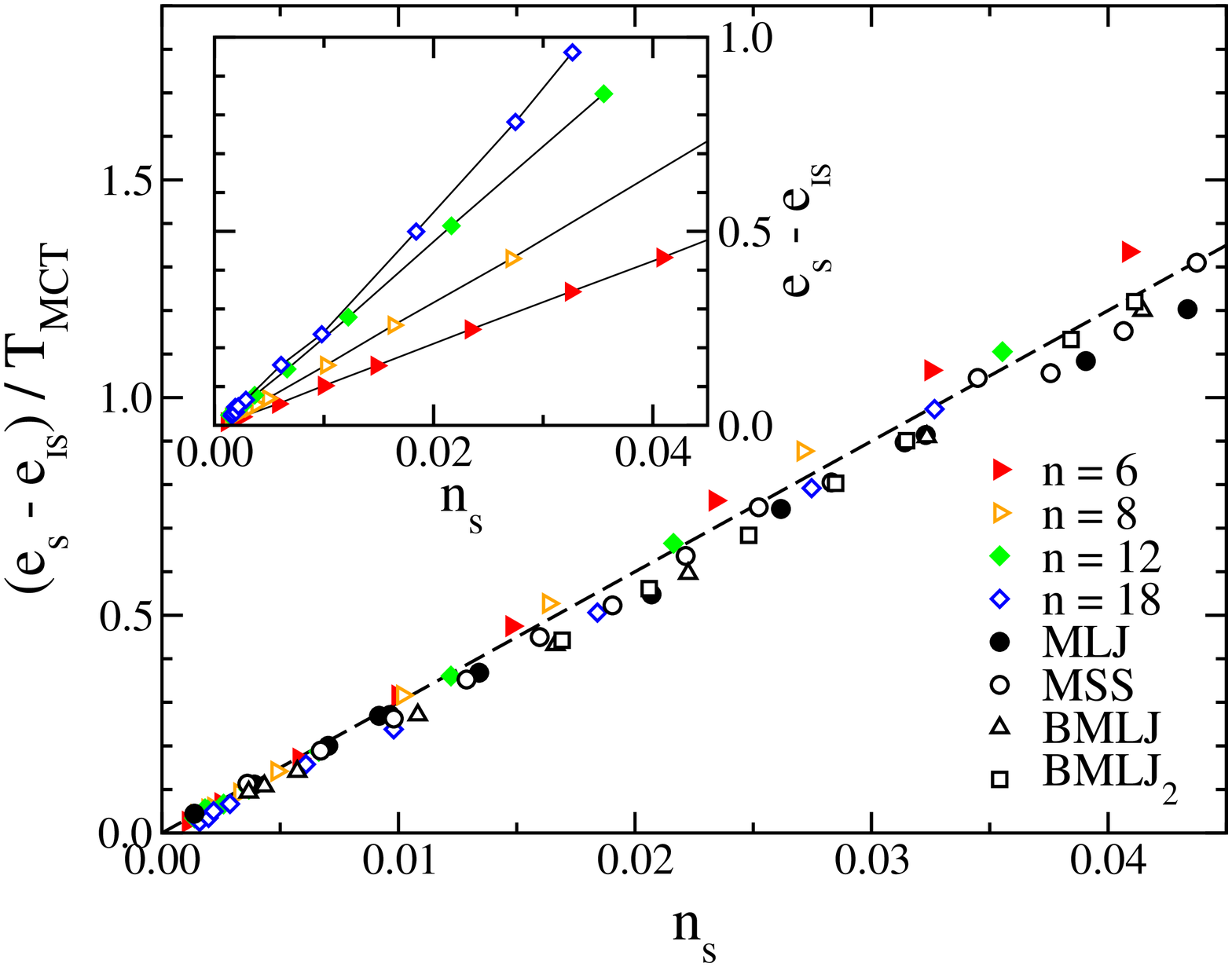}
\caption{
Energy elevation (rescaled by $T_{MCT}$) of saddles from underlying minima 
$(e_s - e_{_{IS}})/T_{MCT}$
against saddle order $n_s$, 
for the systems of this
work (soft spheres with different {\it softness} $n$)
and for LJ-like models
\cite{jcp_ang1}.
The straight line has slope $30$, corresponding to an elementary
energy barrier $\Delta E/T_{MCT}$=$10$.
In the inset the same quantity not rescaled by $T_{MCT}$ for the soft sphere models.
}
\label{fig2}
\end{figure}
This indicates a simple organization of the PEL, with saddles equispaced
in energy above underlying minima, and allows to define an elementary 
energy barrier elevation $\Delta E$ from the slopes $m'$ of the curves:
$\Delta E = m'/3$ (the factor $3$ arises from the definition of $n_s$
as the number of negative eigenvalues of the Hessian 
normalized to the number of degrees of freedom $3N$).
From the inset in the figure one observes that different {\it softness} 
give rise to 
different $\Delta E$ values.
In the main panel of Fig. 2 the energy elevation, 
now scaled by $T_{MCT}$, is plotted vs. $n_s$. 
The data collapse onto a master curve, with 
a mean slope $m=m'/T_{MCT}=30$, indicating that the different soft sphere models
have a similar landscape organization with the same elementary energy barrier
height when expressed in unit of $T_{MCT}$: $\Delta E / T_{MCT} \simeq 10$.
We note, however, a small correlation 
between the slope and the value of $n$, with lower values of $n$ 
associated to higher values of the slope. 
Also reported in Fig. 2 are the data obtained in Ref.~\cite{jcp_ang1}, 
suggesting a wider universality of the relationship $\Delta E \simeq 10 \;T_{MCT}$.

A last observation arises from the analysis of Arrhenius energy barriers
$\Delta E_{Arr}$, obtained from Arrhenius fits $\exp (-\Delta E_{Arr}/T)$
of the low-$T$ diffusivity data (see Ref. \cite{jcp_ang1}).
Expressed in unit of $T_{MCT}$ one obtains, for different {\it softness} $n$,
values of $\Delta E_{Arr}$ in the range $21-24$.
Again this finding is in agreement with the previous observation
for LJ-like models: $\Delta E_{Arr} \simeq 2\; \Delta E$.

In conclusion, the analysis of soft sphere models with different {\it softness}
confirms the results previously obtained for LJ-like models,
supporting the hypothesis of an universal relation controlling 
the structure of the PEL.


\begin{thebibliography}{99}

\bibitem{deb_nature}
P.G.~Debenedetti and F.H.~Stillinger,
Nature {\bf 410}, 259 (2001).

\bibitem{inm}
T.~Keyes, J. Chem. Phys. {\bf 101}, 2921 (1997).

\bibitem{nature1}
S.~Sastry, P.G.~Debenedetti and F.H.~Stillinger,
Nature {\bf 393}, 554 (1998).

\bibitem{heuer}
S.~B\"uchner and A.~Heuer,
Phys. Rev. Lett. {\bf 84}, 2168 (2000).

\bibitem{pel_th}
F.~Sciortino, W.~Kob, and P.~Tartaglia,
Phys. Rev. Lett. {\bf 83}, 3214 (1999).

\bibitem{donati}
C.~Donati, F.~Sciortino and P.~Tartaglia,
Phys. Rev. Lett. {\bf 85}, 1464 (2000).

\bibitem{sad_ag}
L.~Angelani, R.~Di Leonardo, G.~Parisi, and G.~Ruocco,
Phys. Rev. Lett. {\bf 87}, 055502 (2001).

\bibitem{pel_ag}
F.~Sciortino and P.~Tartaglia,
Phys. Rev. Lett. {\bf 86}, 107 (2001).

\bibitem{noi_sad}
L.~Angelani, R.~Di Leonardo, G.~Ruocco, A.~Scala and F.~Sciortino,
Phys. Rev. Lett. {\bf 85}, 5356 (2000);
J. Chem. Phys. {\bf 116}, 10297 (2002);
{\it ibid} {\bf 118}, 5265 (2003).

\bibitem{cavagna}
K.~Broderix, K.K.~Bhattacharya, A.~Cavagna, A.~Zippelius, and I.~Giardina, 
Phys. Rev. Lett. {\bf 85}, 5360 (2000).

\bibitem{chakra}
P.~Shah and C.~Chakravarty,
Phys. Rev. Lett. {\bf 88}, 255501 (2002); 
J. Chem. Phys. {\bf 118}, 2342 (2003).

\bibitem{deb_prl}
M.S.~ Shell, P.G.~Debenedetti, and A.Z.~Panagiotopoulos,
Phys. Rev. Lett. {\bf 92}, 169902 (2004).

\bibitem{jcp_ang1}
L.~Angelani, G.~Ruocco, M.~Sampoli, and F.~Sciortino,
J. Chem. Phys. {\bf 119}, 2120 (2003).

\bibitem{demic}
C.~De Michele, F.~Sciortino, and A.~Coniglio,
cond-mat/0405282.

\bibitem{note}
Redundant $\epsilon$ and $\sigma$ are used in order to emphasize the
similarity with the LJ-binary mixtures \cite{bmlj}.

\bibitem{bmlj}
W.~Kob and H.C.~Andersen,
Phys. Rev. Lett. {\bf 73}, 1376 (1994).

\bibitem{lbfgs}
D.C.~Liu and J.~Nocedal, 
Math. Program. {\bf 45}, 503 (1989).


\bibitem{REM_keyes}
T.~Keyes, J.~Chowdhary, and J.~Kim,
Phys. Rev. E {\bf 66}, 051110 (2002).


\bibitem{REM_note} 
The power-law fitting function is 
$n_s(x)$=$c(x-1)^\gamma$, with $c$=$0.0243(4)$ and $\gamma$=$0.904(1)$.
The REM-like fitting function \cite{REM_keyes} is 
$n_s(x)$=$(b/2)\;[1-erf(1/(2ax))]$ ($erf$ is the error function), 
with $a$=$0.296(5)$ and $b$=$0.215(5)$,
very close to those reported in Ref. \cite{REM_keyes}.

\end{thebibliography}
\end{document}